\documentclass[aps,prl,twocolumn,superscriptaddress]{revtex4}

\usepackage{graphicx}
\usepackage[german,english]{babel}
\newcommand{\ang}{\ensuremath{\rm{\AA}}}

\newcommand{\s}{{\sigma}}

\newcommand{\bea}{\begin{eqnarray}}
\newcommand{\eea}{\end{eqnarray}}
\newcommand{\be}{\begin{equation}}
\newcommand{\ee}{\end{equation}}

\newcommand{\re}{\ensuremath{\rm{e}}}
\newcommand{\ri}{\ensuremath{\rm{i}}}
\newcommand{\rd}{\ensuremath{\rm{d}}}
\newcommand{\p}{\ensuremath{\partial}}

\newcommand{\beqa}{\begin{eqnarray}}
\newcommand{\eeqa}{\end{eqnarray}}

\begin{document}

\title{Midgap states in corrugated graphene: Ab-initio calculations and effective field theory}

\author{T. O. Wehling}
\affiliation{I. Institut f{\"u}r Theoretische Physik, Universit{\"a}t Hamburg, Jungiusstra{\ss}e 9, D-20355 Hamburg, Germany}
\author{A. V.  Balatsky}
\affiliation{Theoretical Division, Los Alamos National Laboratory,
Los Alamos, New Mexico 87545, USA}
\author{A. M. Tsvelik}
\affiliation{Department of Condensed Matter Physics and Materials
Science, Brookhaven National Laboratory, Upton, NY 11973-5000,
USA}
\author{M. I. Katsnelson}
\affiliation{Institute for Molecules and Materials, Radboud
University of Nijmegen, Heijendaalseweg 135, 6525 AJ Nijmegen, The
Netherlands}
\author{A. I. Lichtenstein}
\affiliation{I. Institut f{\"u}r Theoretische Physik, Universit{\"a}t Hamburg, Jungiusstra{\ss}e 9, D-20355 Hamburg, Germany}


\begin{abstract}
We investigate the electronic properties of corrugated graphene and show how rippling-induced
pseudomagnetic fields alter graphene's low-energy electronic properties by
combining first principles calculations with an effective field theory. The
formation of flat bands near the Fermi level corresponding to
pseudo-Landau levels is studied as a function of
the rippling parameters. Quenched and relaxed ripples turn out to be fundamentally different is this respect: It is demonstrated, both numerically and analytically, that annealing of quenched ripples can destroy the flat bands.
\end{abstract}

\date{\today}


\date{\today}

\maketitle

Graphene, i.e. a monolayer of graphite, is the first truly two
dimensional (2D) material available for experiments
\cite{Novoselov_science2004,K.S.Novoselov07262005}. Its low energy
electronic structure resembling 2D Dirac massless fermion dynamics
with the speed of light $c$ being replaced by the Fermi velocity
$v_f\approx c/300$ makes ultrarelativistic physics observable in
this material \cite{kostya2,kim,KatsnelsonRev2007}. These
extraordinary electronic properties are immediately related to the
graphene's 2D honeycomb crystal structure, which was puzzling
itself: Thermal fluctuations in two-dimensional solids, in
principle, should lead to huge displacements of the carbon atoms
from their perfect lattice arrangement and destroy any long range
crystalline order.
As an important step towards solving this puzzle, transmission
electron microscopy \cite{meyer2007} and atomistic
simulations \cite{fasolino} found that free standing graphene sheets are
not perfectly flat but exhibit ripples. In addition to this intrinsic
crumpling, graphene's bonding to a substrate can also introduce
rippling \cite{ElenaStolyarova,Ishigami}. Both, the intrinsic and extrinsic ripples, may be accompanied by
different strengths of in plane strain.


The effect of curvature and strain on
the electronic structure of graphene can be described by effective
gauge fields $A$ acting on the electrons \cite{morozov,morpurgo} -
in close analogy to curvature effects in carbon nanotubes (CNTs) \cite{Kane1997,Lammert2000}.
There, curvature decisively determines the low energy electronic properties, as it can open band gaps: The induced effective
gauge fields shift the Fermi points of the 2D graphene bands away
from the 1D nanotube bands. In graphene, on the contrary, the
Dirac points are similarly shifted by uniform gauge fields but no gap opening is
expected unless a modulated electrostatic potential is present
\cite{guinea-2007}. Rather, the Dirac point shifting leads to
unusually strong electron-phonon interactions \cite{Pisana2007}.

However, the rippling induced gauge fields $A$ are nonuniform and affect
the electrons in graphene like an \textit{effective} magnetic
field ${\cal B}=\nabla\times A$ \cite{morozov}.
Tight-binding (TB) estimations on the effective magnetic field induced
to graphene by rippling found the possibility of
partially flat bands, which are the analog of Landau levels in
real magnetic fields \cite{guinea-2007}. Therefore, these flat
bands are referred to as \textit{pseudo} Landau levels. 
In particular, zero-energy chiral states ($n=0$\, LL) at the
Fermi level should occur in inhomogeneous (real and effective)
magnetic fields, as was pointed out from topological
considerations \cite{kostya2,prokhorova}.
The appearance of ripple-induced mid-gap states can
lead to important consequences: 
The increased DOS at the Fermi level will enhance the tendency to spatial
inhomogeneities \cite{guinea-2007} as well as lead to strong resonant electron scattering
\cite{KatsnelsonRev2007,katsnelson0706,stauber-2007}. Knowing the conditions under which midgap states occur is therefore essential to resolve the debate on electron scattering in graphene.

So far, the predictions on midgap states have been rather qualitative involving
adjustable parameters, like hopping matrix elements, their change
with strain and curvature as well as they neglected
rehybridization effects of the $\pi$ and $\sigma$ bands. It is not
clear {\it a priori} how essential these effects can be. For
example, taking into account next-nearest-neighbor hopping leads
to a {\it scalar} electrostatic potential induced by the ripples
\cite{castronetokim} which can cause opening a gap around the Fermi level \cite{guinea-2007}. Only based on
complete first-principles calculations one can judge how important
the corrections to the simplest nearest-neighbor
TB model are.

In this letter, we present full potential density functional theory (DFT) studies of quenched and annealed graphene ripples. We find flat bands very close to the Dirac point for quenched ripples, whereas in annealed ripples these midgap states turn out to be suppressed. Based on a nearest neighbor TB model, we extend the low energy effective field theory description of graphene to include this relaxation effect. The qualitative agreement of our effective field theory with DFT justifies this nearest neighbor TB based theory for describing corrugated graphene.

\begin{figure}
\centering
\begin{minipage}{.98\linewidth}
 \resizebox{80mm}{!}{\includegraphics{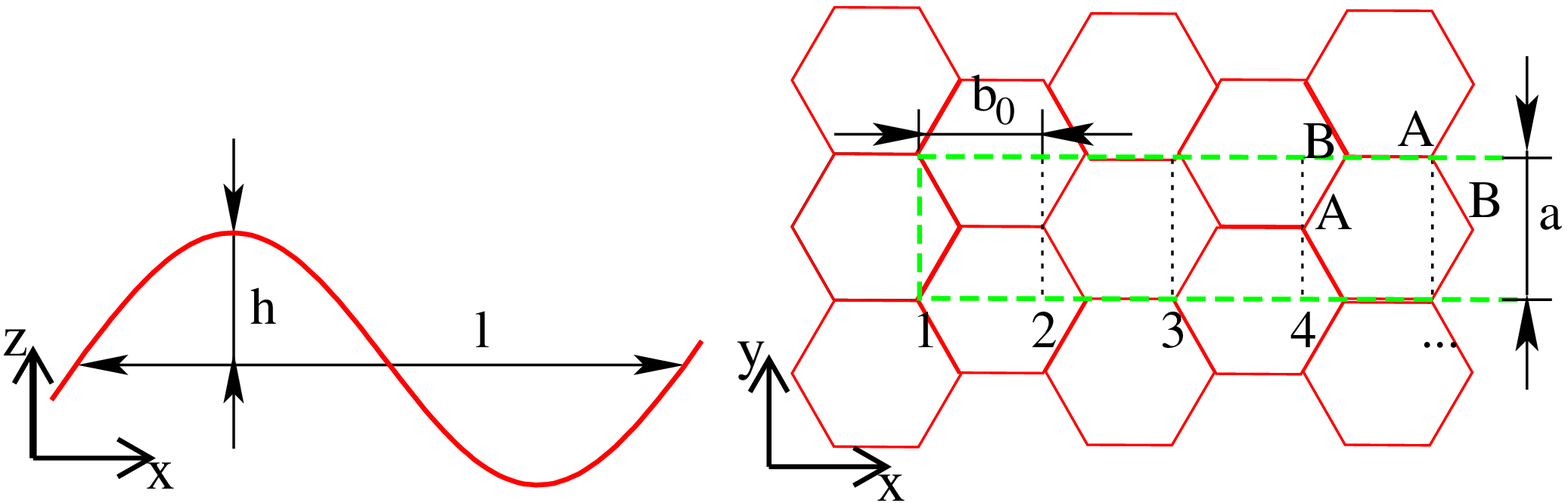}}\vspace{.3cm}
\end{minipage}
\begin{minipage}{.98\linewidth}
\resizebox{80mm}{!}{\includegraphics{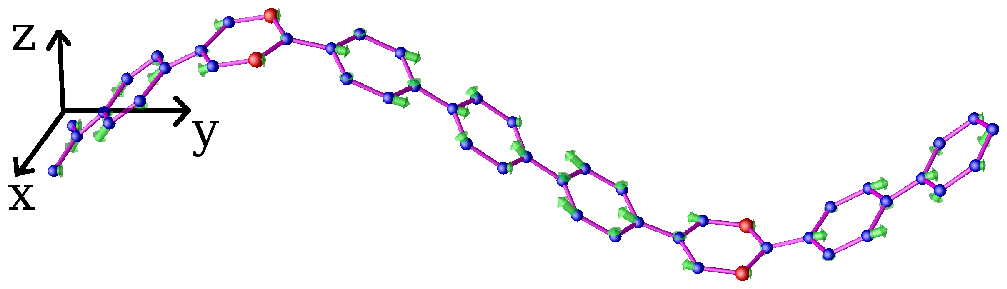}}
\end{minipage}

\caption{\label{fig:setup}(Colour online) Upper panel: Schematic top and side
view of the sinusoidal graphene ripples. The rippling period is denoted by
$l$ and their amplitude by $h$. Lower panel: Perspective view of a relaxed graphene ripple. The green arrows show the displacement of the atoms during the relaxation. To enforce a constant rippling amplitude the vertical position of the atoms marked as big red dots has been fixed.}
\end{figure}

In our DFT calculations, quenched ripples are modelled as sinusoidal graphene ripples with height field $h(x,y)=h_0\sin(qx)$ (Fig. \ref{fig:setup}, upper panel).
While the effective gauge field $A$ is caused by strain and curvature,
the strain in the ripple can be significantly reduced by allowing
all atoms and the supercell shape to relax with the constraint of fixed rippling amplitude $h_0$. As starting point for the relaxed cell shape, we shortened the supercell in rippling direction such that the arc length of one sinusoidal graphene ripple period coincides with the equilibrium length of the same supercell for flat graphene. Then, standard relaxation of the atomic positions and the cell shape under the rippling constraint from above has been performed.
This relaxation leads to our model of annealed graphene ripples, where "annealed" means that in these ripples possible external sources for energy barriers preventing the ripples from relaxation have been removed. Such sources might be impurities or bonding to a substrate with lattice mismatch.\cite{Roehrl,lauffer}

In all of our calculations we employ the generalized gradient approximation (GGA)
\cite{Perdew:PW91} to DFT for supercells containing up to
160 carbon atoms. The resulting Kohn-Sham problems are
solved with the Vienna Ab Initio Simulation
(VASP) \cite{Kresse:PP_VASP} package by expanding the electronic
bands into projector augmented waves (PAW)
\cite{Kresse:PAW_VASP,Bloechl:PAW1994}. Plane wave cut-offs of
$500$\,eV for band-structure and $875$\,eV for the relaxations and
total energy calculations were used. For the total energy calculations, the Brillouin
zone integrations were performed with $0.1$ eV Gaussian smearing
on k-meshes denser than $24\times 24$ when
folded back to the graphene first Brillouin zone, whereas for
relaxations and input charge densities for band
structure calculations k-meshes diluted by a factor of 2 turned
out to be sufficient.

\begin{figure}
\centering
\resizebox{70mm}{!}{\includegraphics{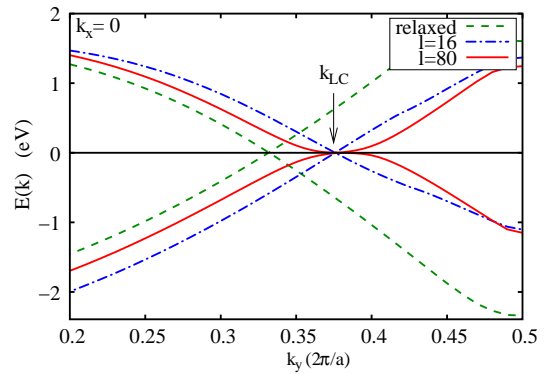}}
\caption{\label{fig:bandstruct}(Colour online) Band structure of
corrugated graphene sheets: The highest valence and lowest
conduction bands along the $k_y$
direction (perpendicular to the rippling direction) are shown in main panel for different armchair ripples. Dash dotted and solid lines: Unrelaxed sinusoidal ripples with fixed $h_0/l=1\ang/4b_0$ ratio for $l= 16,\,80 b_0$. Dashed lines: A ripple relaxed with the constraint $h_0=4\ang$ and a rippling period of 16 graphene unit cells. 
The energy required to create the quenched ripples considered, here, is $0.5$\, eV$/$atom. For the relaxed ripple this energy is $0.03$\, eV$/$atom}
\end{figure}

Firstly, the occurrence of the $n=0$\, LL in quenched ripples as function of the ripple length $l$
and $h_0/l$ has been studied, in this way. For $h_0/l=1\ang/4b_0$ with $b_0=\sqrt{3}a_0/2$ and $a_0=2.465\ang$ being the graphene lattice constant, prominent changes in the high valance and the low conduction bands occur (Fig. \ref{fig:bandstruct}) depending on
the rippling period $l$.
The shorter ripple ($l=16b_0$) exhibits electron
dispersion resembling massless particles with the Dirac point
shifted from $k_y=2\pi/3a_0$ to $k_y\approx 0.375(2\pi/a_0)$, where $k_y$ is the crystal momentum perpendicular to the rippling direction, $x$.
However, the bands of the $l=80\,b_0$ ripple are flat near the Fermi level and
exhibit all characteristics of the $n=0$\, LL of Dirac fermions: They are chiral, that is, fully sublattice
polarized and localized in regions of maximum effective magnetic field ${\cal B}$.

\begin{figure}
\centering

     \resizebox{70mm}{!}{\includegraphics{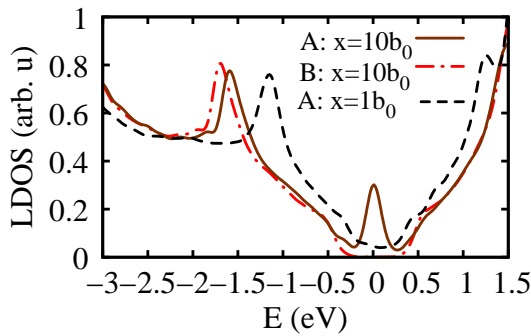}}

\caption{\label{fig:localWF}(Colour online) 
The local density of states (LDOS) inside
the cells at $x=1b_0$ (low eff. field) and at $x=10b_0$ (high field region). For the
low field region, the LDOS is the same in both sublattices (only
sublattice A plotted is here, dashed line), whereas in the high
field region the LDOS in sublattice A (solid) and B (dash-dotted)
differ significantly.}
\end{figure}

This manifests in the local density of states (LDOS) (see Fig. \ref{fig:localWF}) as follows.  For a ripple of the form $h(x,y)=h_0\sin(qx)$ with $q=2\pi/l$ and $l$ sufficiently large, the effective gauge field is $A\sim(hq\cos(qx))^2$ and accordingly
${\cal B}\sim h^2q^3\sin(2qx)$. Thus, for $l=80b_0$ the absolute value of the
effective field is maximum around $x=10,30,50,70b_0$. 
In this high field region (see Fig. \ref{fig:localWF}), the spectrum is gapped around
the Fermi level ($E=0$) in one sublattice (here B) but exhibits a
mid-gap peak in the other sublattice. This will lead to sublattice
stripes in low bias STM topographs with sublattice A bright and B
dark, which is very reminiscent of midgap impurity states \cite{wehling:125425}.
In the low effective field region, the LDOS recovers the pseudogap
shape typical for flat graphene and the two sublattices appear
fully equivalent again.

The band structures of the quenched ripples shown in Fig. \ref{fig:bandstruct} allow to estimate the strength of the involved pseudomagnetic fields: Given the shift of the Dirac point of $\delta k_y=0.042(2\pi/a_0)$ away from the flat graphene value, we obtain the average gauge field $A_0=\hbar\delta k_y/e=2.8\cdot 10^3$\,T$a_0$. Due to the sinusoidal shape of the ripple the amplitude sinusoidal pseudomagnetic field is $\mathcal{B}_0=A_0q\approx 250$T for $q=2\pi/80 b_0$. Due to $\mathcal{B}=\mathcal{B}_0\sin(2qx)$ a pseudo landau level wave function should be localized on a length less than $l/4$ corresponding to an area of $l^2/16=18$\,nm$^2$. (See also Ref. \cite{guinea-2007}.) 250T correspond to approx. 2 (pseudo)magnetic flux quanta per 18nm$^2$.

The ripple under consideration has maximum local strain of $24\%$, which is more than the average strain of approx. $1\%$ measured in epitaxial graphene\cite{Roehrl} or $4\%$ found in other nanosized epitaxial materials\cite{Brune}. Scaling down the pseudo magnetic field to these strain values yields 10T and 40T, respectively.


So far, we considered quenched ripples, but the pseudomagnetic field is sensitive not only to flexural deformations but also to in-plane distortions. In general, ripples will be accompanied by in-plane distortions: For the quenched ripple of length $l=16b_0$ with $h/l=1\ang/4b_0$, we relaxed the atomic positions and the supercell shape of the ripple with the constraint of fixed rippling height. This ''annealed'' ripple is still sinusoidal (Fig. \ref{fig:setup}, lower panel) but with all nearest neighbor bond lengths being equal after relaxation.
During this relaxation process the effective gauge field decreases as can be seen from the band structure of the relaxed ripple in Fig. \ref{fig:bandstruct}. The Dirac point for the relaxed ripple is at $k_y\approx 0.332(2\pi/a_0)$, which is by a factor of 40 closer the flat graphene value of $k_y= 1/3(2\pi/a_0)$ than the Dirac point of the quenched ripples. This corresponds to a decrease of the average effective gauge field the same factor. This suppression of effective gauge and pseudomagentic fields in the annealed ripple can be understood in terms of the following model:

Consider the nearest neighbor
TB Hamiltonian \bea H = \sum_k
(A^+_k, B^+_k)\left(
\begin{array}{cc}
0 & t_{nn}(k)\\
\bar t_{nn}(k) & 0
\end{array}
\right) \left(
\begin{array}{c}
A_k\\
B_k
\end{array}
\right), \label{H0} \eea where $t_{nn}$ are the hopping matrix elements and $A_k$ ($B_k$) the Fermi operators of electrons in sublattice A (B) with crystal momentum $k$. Close to
$\pm {\bf K} = (0,\mp 4\pi/3\sqrt{3})$, in the corner of the Brillouin
zone, the dispersion for the undeformed lattice vanishes linearly and a continuum theory
describing low energy electronic states can be defined. As in Ref. \cite{foster}, we introduce
a pair of two-dimensional spinors $ \Psi_1 = \left(
\begin{array}{c}
 A_K\\
B_K
\end{array}
\right), ~~ \Psi_2 = \left(
\begin{array}{c}
B_{-K}\\
-A_{-K}
\end{array}
\right)$
describing electronic wave packets
centered at ${\bf K}$ and $-{\bf K}$ point, respectively, and
expand the hopping integrals. Near the ${\bf K}$ point the
undeformed hopping integral is 
\begin{equation}t_{nn}({\bf k})\approx \frac{3t_0a_0}{2}(q_y - \ri q_x) \rightarrow 2v_0\bar\p,\end{equation}
where $q = k -K$, $\bar\p = \frac{1}{2}(\p_x +\ri\p_y)$ and $v_0 = 3t_0\tilde a_0/2$ is the Fermi velocity, involving explicitly the nearest-neighbor spacing $\tilde a_0=a_0/\sqrt{3}$. Slow lattice deformations with the deformation tensor field 
\bea u_{ab} = \frac{1}{2}[\p_a u_b + \p_b u_a] + \p_a h\p_b h \eea
with $a,b\in\{x,y\}$ lead to changes in the hopping integral \bea
&& \delta t({\bf Q})/[\frac{\p t_0}{\p R}] =\sum_{\mu}\re^{\ri {\bf Q}{\bf e}_{\mu}}e_{\mu}^a\frac{\p u^a}{\p x^b} e_{\mu}^b 
= 3u_{zz} \eea where
$e_1 = (1,0), e_2 = (1/2,\sqrt 3/2), e_3 = (1/2,-\sqrt 3/2)$ are
vectors connecting nearest neighbor sites and $z = x + i y$. Thus the nearest neighbor hopping
matrix element near the point ${\bf K}$ is now \bea
2v_0\left(\bar\p - \gamma u_{zz}\right) \label{matr} \eea where
$\gamma = \p\ln t_0/\p R$ plays a role of the charge, and  $
u_{zz}$ can be interpreted as a vector potential: \bea \bar A^3 =
u_{zz}, A^3 = A_z^3 \equiv A_x^3 - \ri A_y^3 = u_{\bar{z}\bar{z}}
\label{Au} \eea 

 We have considered smooth lattice deformations and established that they generate an Abelian vector potential having opposite signs for different valleys. It is worth noticing that this vector potential constitutes a part of the non-Abelian field whose other noncommuting components $A^{1,2}$ are generated by abrupt changes of the nearest neighbor hopping integrals. The complete low energy Hamiltonian is \cite{foster} \bea H =
\Psi^+\left\{I\otimes I A_0 + v_0\s_{\mu}\otimes[-\ri  I\p_{\mu} +
\gamma\tau^a A_{\mu}^a]\right\}\Psi \label{2D} \eea where the
Pauli matrices $\s_{\mu}, \mu = x,y$ act in the spinor space and
matrices $\tau^a, a =1,2,3$ act on the valley index of the spinor
$\Psi^+=(\Psi^+_1,\Psi^+_2)$. 


The zero energy wave functions for the vector potential
deformations can be found analytically for $A_0 =0$. For
simplicity we will restrict ourselves to the case of slow
deformations when $A^{1,2}=0$. Then  the zero energy wave
functions (if they exist) are expressed in terms of
pseudo-magnetic field \bea {\cal B} = \ri(\p A^3 - \bar\p\bar A^3)
\label{B} \eea
 For ${\cal B} > 0$ we have:
\bea A_{K} = A_{-K} = z^n\exp\left[- \frac{\gamma}{\p\bar\p}{\cal
B}\right], ~~B_{K} = B_{-K} =0 \label{psi} \eea and for ${\cal B}
< 0$ one has to interchange $A_{K}, A_{-K}$ with $B_{K}, B_{-K}$
and $z$ with $\bar z$. The power $n$ ranges from 0 to the integer
part of the pseudomagnetic flux. As many zero modes exist.

The expression for ${\cal B}$ and for the deformation tensor
changes drastically depending on whether the elastic energy is at
its minimum or not. The expression for the elastic energy density
of a smooth surface compatible with the $C_3$ symmetry is given by
\cite{nelsonbook}
\bea  E&=& (\lambda +
\mu)\left[\frac{1}{2}(\bar\p u + \p\bar u) + \p h\bar\p h\right]^2\nonumber\\
&&+\mu\left[\p u + (\p h)^2\right]\left[\bar\p\bar u + (\bar\p h)^2\right] + \frac{K}{2}(\nabla^2h)^2 \nonumber\\
&=& \mu A\bar A + (\lambda +\mu)\left[\frac{1}{2\p\bar\p}({\bar\p}^2 A + \p^2\bar A)+\frac{1}{\p\bar\p}{\cal R}[h]\right]^2\nonumber\\ &&+ \frac{K}{2}(\nabla^2h)^2
\label{elast} \eea
where $ {\cal R} = [\p^2 h{\bar\p}^2h -(\p\bar\p h)^2]$
is the Gaussian curvature of
the surface. If the membrane is at equilibrium, the elastic energy
(\ref{elast}) is at its minimum  and the vector potential is
given by \bea A^3 = -\frac{(\lambda + \mu)}{(\lambda +
2\mu)}\frac{\p^2}{\p\bar\p}{\cal R}. \label{eqn:gauge_relaxed}\eea  
${\cal R}$ is the Jacobian of the coordinate transformation 
$ \xi = \p h, ~~ \bar\xi = \bar\p h$:\bea \p^2 h{\bar\p}^2h - (\p\bar\p h)^2 =
\frac{\p\left(\xi,\bar\xi\right)}{\p\left( z, \bar
z\right)} \eea  Thus, ${\cal R}$ and,
as a consequence,  $A^3, \bar A^3$ as well as $A_0$ vanish on any
configuration of $h$ which depends on just one Cartesian
coordinate. This includes all plane wave configurations, in particular those studied by DFT in this letter.

Substituting Eqn. (\ref{eqn:gauge_relaxed}) into
(\ref{B}) we get \begin{equation} {\cal B} = \frac{\ri(\lambda + \mu)}{\lambda
+ 2\mu}\frac{\p^3 - {\bar\p}^3}{(\p\bar\p)^2}[\p^2 h{\bar\p}^2h -
(\p\bar\p h)^2], \end{equation} which is  $C_3$ symmetric, as it
must be. This $C_3$ symmetry of ${\cal B}$ for a relaxed
membrane leads to important qualitative differences with a real
magnetic field. A local vortex of magnetic field with flux $N$
carries $N$ normalizable zero modes, but  to create such a flux by
deforming a membrane is not possible. An analogue of a magnetic
vortex is a point-like defect $h = h(|{\bf r -a}|)$ carrying local Gaussian curvature
${\cal R} = {\cal R}(|{\bf r-a}|)$. Then the argument in wave function (\ref{psi}) is $
\frac{\gamma}{\p\bar\p}{\cal B} = \gamma\sin(3\phi)b(|r -a|)$, where
$b(r) = \int \frac{\rd k}{(2\pi)k^2}J_3(kr){\cal R}_k $ and
$\phi$ is the angle with respect to the crystalline axis. Since
${\cal R}_k$ is constant at small $k$, at distances larger than
the size of the defect $b(r) \sim r$. Such wave function is not
normalizable. On the other hand the wave function with two defects
with curvature of the opposite sign will be normalizable for the
entire volume (as a plane wave).  Such  state is non-degenerate,
that is we have $n=0$ in  (\ref{psi}).
So, if we allow the graphene membrane to relax there are only nondegenerate zero modes, which only exist
for membrane configurations with nonzero Gaussian curvature.
Therefore, the effect of relaxations on the electronic properties of the graphene ripples is qualitatively the same for 1D and 2D ripples. The \textit{degenerate} zero modes are suppressed in relaxed ripples and no significant midgap peak in the total DOS of 2D ripples is expected to occur.

In conclusion, our results demonstrate an essential difference between
quenched and annealed ripple structure. If the system is allowed
to relax to its minimum of elastic energy for a given $h(x,y)$
profile it decreases drastically the amplitude of pseudo-magnetic
fields and can lead to disappearance of the mid-gap states. It may
be an important statement in light of the hypothesis
\cite{katsnelson0706} that \textit{quenched} ripples are the main
source of scattering in graphene and the very recent observation
that annealing of a freely hanged graphene membrane can increase
drastically its mobility \cite{freehanged}: Upon annealing, impurities causing energy barriers that prevent the suspended graphene from relaxation might be removed. This issue requires further investigations. Our ab-initio calculations reveal a perfect particle-hole symmetry at low energies and justify the nearest neighbor hopping based field theory for describing the electronic properties of graphene ripples.

The authors are thankful to A. Geim, K. Novoselov, and I. Aleiner for inspiring discussions. This work
was supported by SFB 668 (Germany), FOM (The Netherlands) and DOE
at Los Alamos. The authors acknowledge computer time from LANL (USA) and HLRN (Germany). T.O.W. is grateful to LANL for hospitality during the
visit, when the ideas presented in this work were set off. A.M.T.
acknowledges the support from US DOE under contract number DE-AC02
-98 CH 10886.
\bibliography{ripples_s}
\end{document}